\documentclass[twoside]{article}
\usepackage[latin1]{inputenc}
\usepackage{t1enc}
\usepackage{a4}
\usepackage{tabularx}
\usepackage{epsf}
\usepackage{amsmath,amssymb,amsfonts,mathrsfs,latexsym} 
\usepackage{graphicx}
\usepackage{makeidx}
\makeindex
\newcommand{\TT}{\mathrm{TT}}

\newcommand{\Ibar}{{\cal I} \kern -0.5em\raise 0.25ex \hbox{--}}
\newcommand{\be}{\begin{equation}}
\newcommand{\ee}{\end{equation}}
\begin{document}

\setcounter{figure}{0}
\setcounter{section}{0}
\setcounter{equation}{0}

\begin{center}
{\Large\bf
Gravitational Wave Astronomy
}\\[0.7cm]

Kostas D. Kokkotas \\[0.17cm]
Theoretical Astrophysics, 
Auf der Morgenstelle 10, \\ Eberhard Karls University of T\"ubingen, T\"ubingen 72076, Germany
\footnote{Reviews in Modern Astrophysics, Vol 20, "Cosmic Matter", WILEY-VCH, Edited by S. R\"oser (2008)}\\
\end{center}

\vspace{0.5cm}

\begin{abstract}
\noindent{\it

As several large scale interferometers are beginning to take 
data at sensitivities where astrophysical sources are predicted, 
the direct detection of gravitational waves may well be imminent. 
This would  open the gravitational-wave
window to our Universe, and should lead to a much improved
understanding of the most violent processes imaginable; 
the formation of black holes and neutron stars following core
collapse supernovae and the merger of compact objects at the end of 
binary inspiral.

}
\end{abstract}

\section{Introduction}
Gravitational waves are ripples of spacetime generated as masses are
accelerated.
It is one of the central predictions of Einstein's general theory of relativity 
but despite decades of effort these ripples in spacetime have still not 
been observed directly. Yet we have strong indirect evidence for their 
existence from the excellent agreement between the observed 
inspiral rate of the binary pulsar PSR1913+16 and the theoretical
prediction (better than 1\% 
in the phase evolution). 
This provides confidence in the theory and suggests that 
``gravitational-wave astronomy'' should be viewed as a
serious proposition. 
This new window onto  universe
will complement our view of the  cosmos  and will  help  us unveil
the fabric of spacetime around  black-holes, observe directly the
formation of black holes or  the merging  of  binary systems
consisting  of  black holes  or neutron  stars,  search for
rapidly spinning neutron  stars, dig deep into the very early
moments of the origin of the universe, and look at the very center
of the galaxies where  supermassive black holes weighing millions
of  solar masses  are  hidden.   Secondly, detecting
gravitational  waves is important for our understanding of the
fundamental   laws   of  physics;   the proof   that gravitational
waves exist will verify a fundamental 90-year-old prediction of
general relativity. Also, by comparing the arrival times of light
and gravitational waves from,  e.g., supernovae, Einstein's
prediction   that    light    and gravitational waves  travel at
the  same  speed  could  be checked.  Finally, we  could  verify
that  they  have  the polarization predicted by general
relativity.

These expectations follow from the comparison between 
gravitational and electromagnetic waves. That is : a)
 While electromagnetic waves are radiated by individual particles, 
gravitational waves are due to non-spherical bulk motion of matter. 
In essence, this means that the information carried by electromagnetic 
waves is stochastic in nature, while the gravitational waves provide 
insights into coherent mass currents.   
b) The electromagnetic waves will have been 
scattered many times. In contrast, gravitational
waves interact weakly with matter
and arrive at the Earth in pristine condition. 
This means that gravitational waves can be used to probe regions
of space that are opaque to electromagnetic waves. Unfortunately, this weak interaction with matter also makes the detection of gravitational waves an extremely hard task.
c) Standard astronomy is based on deep
imaging of small fields of view, while gravitational-wave 
detectors cover virtually the entire sky.  
d) The wavelength of the electromagnetic radiation is  smaller
than the size of the emitter, while  the
wavelength of a gravitational wave is usually 
larger than the size of the source. This means that we cannot 
use gravitational-wave data to create an image of the source. 
In fact, gravitational-wave observations are more like 
audio than visual.

\section{Gravitational wave primer}

The aim of the first part of our contribution is to provide a 
condensed text-book level introduction to  gravitational waves.
Although in no sense complete, this description should prepare the 
reader for the discussion of high-frequency sources which follows.

The first aspect of gravitational waves that we need to appreciate is their
\emph{tidal} nature. This is important because it implies that 
we need to monitor, with extreme precision, the relative motion of test masses or the periodic (tidal) deformations of extended bodies.
A gravitational wave,
propagating   in   a  flat  spacetime,  generates   periodic
distortions, which can be described in terms of the  Riemann
tensor which measures the curvature of the spacetime.
 In linearized theory ($h_{\mu\nu}\ll g_{\mu\nu}$) the Riemann tensor  takes  the
following gauge-independent form:
\begin{equation}
R_{\kappa\lambda\mu\nu} =\frac{1}{
2}\left(\partial_{\nu\kappa}h_{\lambda\mu} +
\partial_{\lambda\mu}h_{\kappa\nu} -
\partial_{\kappa\mu}h_{\lambda\nu} -
\partial_{\lambda\nu}h_{\kappa\mu}\right),
\end{equation}
which is considerably simplified by choosing the so called transverse and traceless gauge or \emph{TT gauge}:
\begin{equation}
R_{j0k0}^{\rm TT}=-\frac{1}{2}\frac{\partial^2} { \partial
t^2}h_{jk}^{\rm TT} \approx \frac{\partial^2 \Phi} { \partial x^j \partial
x^k},
 \quad j,k=1,2,3. \label{eq17a}
\end{equation}
where $h_{jk}^{\rm TT}$ is the gravitational wave field in the TT-gauge and $\Phi$  describes the gravitational potential  in Newtonian
theory.  The  Riemann  tensor    is a pure
geometrical  object, but in general relativity has a  simple  physical
interpretation:  it is the tidal force field  and  describes the
relative  acceleration between two  particles  in  free fall.  If
we  assume  two  particles  moving  freely  along geodesics of a
curved spacetime with coordinates $x^\mu(\tau)$ and
$x^\mu(\tau)+\delta x^\mu(\tau)$ (for a given value of the proper time
$\tau$, $\delta x^{\mu}(\tau)$ is the displacement vector connecting
the two events) it can be shown that,  in the case of slowly
moving particles,
\begin{equation}
\frac{d^2\delta x^k}{ d t^2}\approx - {{R^k} _{0j0}}^{ {\rm TT}} \delta x^j .
\label{eq16}
\end{equation}
This  is  a simplified  form of the  equation  of  {\em geodesic
deviation}. Hence, the tidal force acting on a particle is:
\begin{equation}
f^k\approx -m {R^k}_{0j0}\delta x^j, \label{eq17}
\end{equation}
where $m$ is the mass of the particle. Equation (\ref{eq17})
corresponds to the  standard  Newtonian relation for  the  tidal
force acting on a particle in a field $\Phi$.
Then equation (\ref{eq17}) integrates to 
\be
\delta x_j = \frac{1}{2}  h_{jk}^{\rm TT}  x_{0}^k \quad \mbox{or} \quad
h \approx \frac{ \Delta L} { L} \, ,
\ee
where $h$ is the dimensionless gravitational-wave strain.

\begin{figure}
 \centerline{\includegraphics[width=8.5cm]{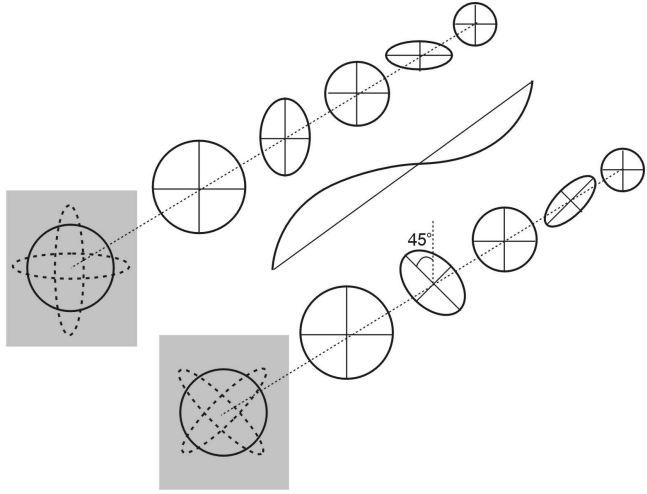}}
\caption{The effects of a gravitational wave travelling
perpendicular the plane of a circular ring of particles, is
sketched as a series of snapshots. The deformations due the two
polarizations  $h_+$ and $h_\times$ are shown.}
\label{Fig1}\end{figure}
Let us now assume that the waves propagate in the $z$-direction, i.e.
that we have $h_{jk}=h_{jk}(t-z)$. Then one can show that 
we have only two independent components;
\be
h_+ = h_{xx}^{\TT} = -  h_{yy}^{\TT} \, , \qquad
h_\times = h_{xy}^{\TT} =   h_{yx}^{\TT}
\ee
What effect does $h_+$ have on matter? 
Consider a particle initially located at $(x_0,y_0)$ and 
let $h_\times=0$ to find that 
\begin{equation}
\delta x = \frac{1}{2} h_+ x_0 \quad \mbox{and} \quad
\delta y = - \frac{1}{2} h_+ y_0 \, .
\end{equation}
That is, if $h_+$ varies periodically then an object 
will first experience a stretch in the $x$-direction accompanied
by a squeeze in the $y$-direction. One half-cycle later, the 
squeeze is in the $x$-direction and the stretch in the $y$-direction. 
It is straightforward to show that
the effect of  $h_\times$ is  the same, but rotated by 45 degrees.
This is illustrated in Fig.~\ref{Fig1}.
A general wave will be a linear combination of the two polarisations. 

Up to this point we have shown the effect of propagating spacetime deformations (we called them gravitational waves) on two nearby particles. But we have not yet done the connection to Einstein's theory. Thus we will describe with a tensor $h_{\mu\nu}$ the variations of a flat spacetime from flateness, that is we will describe the spacetime with the metric
$g_{\alpha\beta} = \eta_{\alpha\beta} + h_{\alpha\beta}$. Then after some algebra it can be shown that Einstein's equations will be reduced to the following form:
\begin{equation}
\left(-\frac{\partial^2}{ \partial t^2}+\nabla^2 \right){\tilde
h}^{\mu\nu}\equiv \partial_\lambda \partial^\lambda {\tilde
h}^{\mu\nu}=0 \quad \mbox{with} \quad {\tilde h}_{\mu \nu}\equiv h_{\mu \nu} - \frac{1}{2} \eta_{\mu\nu}h^\alpha_\alpha \label{eq6}
\end{equation}
where we have used a specific gauge choice  $\partial_\mu {\tilde
h}^{\mu\nu}=0$, known as {\em Hilbert's gauge condition}
(equivalent to the Lorentz gauge condition of electromagnetism).

 The  simplest solution to the wave equation  (\ref{eq6})  is  a
plane wave solution of the form
\begin{equation}
{\tilde h}^{\mu\nu}=A^{\mu\nu}e^{i k_\alpha x^\alpha},
\end{equation}
where $A^{\mu\nu}$ is   a  constant  symmetric  tensor,  the {\em
polarization tensor},  in  which information about the amplitude
and  the polarization  of the waves is encoded, while $k_\alpha$
is a constant vector,  the  {\em wave  vector}, that determines
the propagation direction  of  the  wave  and  its  frequency.  In
physical applications  we will use only the real part  of  the
above wave solution. 

It is customary to write the gravitational wave solution
in the TT gauge as $h_{\mu\nu}^{TT}$. That $A_{\mu\nu}$ has only
two independent components means that a gravitational wave is
completely described by two dimensionless amplitudes, $h_+$ and
$h_\times$, say. If, for example, we assume a wave propagating
along the z-direction, then the amplitude $A^{\mu\nu}$ can be
written as
\begin{equation}
A^{\mu\nu}=h_+ \epsilon_+^{\mu\nu}+
h_\times\epsilon_\times^{\mu\nu} \label{eq11}
\end{equation}
where $\epsilon_+^{\mu\nu}$ and  $\epsilon_\times^{\mu\nu}$ are
the so-called {\em unit polarization tensors} defined by
\begin{equation}
\epsilon_{+}^{\mu \nu} \equiv \left(
\begin{array}{cccc}
   0 & 0 &  0 & 0\\
   0 & 1 &  0 & 0\\
   0 & 0 & -1 & 0\\
   0 & 0 &  0 & 0\\
\end{array}
\right) \qquad \qquad \epsilon_{\times}^{\mu \nu} \equiv \left(
\begin{array}{cccc}
   0 & 0 &  0 & 0\\
   0 & 0 &  1 & 0\\
   0 & 1 &  0 & 0\\
   0 & 0 &  0 & 0\\
\end{array}
\right) \, . \label{eq12}
\end{equation}
Finally, we should point out that in the TT-gauge there is no difference between $h_{\mu\nu}$(the perturbation of the metric) and ${\tilde h}_{\mu\nu}$ (the gravitational
field).

\subsection {Gravitational wave properties}     

Gravitational waves, once they are generated, propagate
almost unimpeded.  Indeed, it has been proven that they  are even
harder  to  stop than neutrinos! The only  significant change they
suffer as they propagate is  the  decrease  in amplitude while
they travel away from their source, and  the {\em redshift} they
feel (cosmological, gravitational or Doppler) as is the case for
electromagnetic waves.

     There  are other effects that marginally influence  the
gravitational   waveforms.  For  instance   {\em absorption}   by
interstellar or intergalactic matter intervening between the
observer  and the source, which is extremely weak (actually, the
extremely  weak  coupling of gravitational  waves  with matter is
the main reason that gravitational waves have  not been observed).
{\em Scattering}   and   {\em dispersion}   of gravitational waves
are also practically unimportant, although they may  have been
important during the early phases of  the universe (this   is also
true for   the   absorption). Gravitational waves can be {\em
focused} by strong  gravitational fields  and  also can be {\em
diffracted}, exactly as  it  happens with the electromagnetic
waves.

\subsection{Energy flux carried by gravitational waves}

     Gravitational   waves  carry   energy   and   cause   a
deformation  of  spacetime.  The  stress-energy  carried  by
gravitational waves cannot be localized within a wavelength.
Instead,  one can say that a certain amount of stress-energy is
contained  in a region of the space which  extends  over several
wavelengths. It can be proven that in the TT  gauge of
linearized  theory  the  stress-energy  tensor   of   a
gravitational wave (in analogy with the stress-energy tensor of a
perfect fluid) is given by
\begin{equation}
t_{\mu\nu}^{GW}=\frac{1}{32 \pi} \langle \partial_\mu
h_{ij}^{TT} \cdot \partial_\nu h_{ij}^{TT} \rangle .
\label{eq24}
\end{equation}
where  the  angular brackets are used to indicate  averaging over
several wavelengths. For  the special case of a plane wave
propagating in the  $z$ direction,  which  we considered earlier,
the  stress-energy tensor  has only three non-zero components,
which  take  the simple form
\begin{equation}
t_{00}^{GW}=\frac{ t_{zz}^{GW} }{ c^2}= - \frac{t_{0z}^{GW} }{c}
                    =\frac{1}{ 32 \pi}\frac{c^2}{G}\omega^2 \left(
h_+^2+h_\times^2\right), \label{eq25}
\end{equation}
where $t_{00}^{GW}$ is the energy density, $t_{zz}^{GW}$ is the
momentum flux and $t_{0z}^{GW}$  the energy  flow  along the $z$
direction per unit area  and  unit time (for  practical reasons we
have restored  the  normal units). The energy flux has all the
properties  one  would anticipate by analogy with electromagnetic
waves: (a) it  is conserved (the amplitude dies out as $1/r$, the
flux as $1/r^2$), (b) it can  be  absorbed by  detectors, and (c)
it can generate curvature like  any other  energy source  in
Einstein's formulation  of relativity. 

The definition of the energy flux by equation (\ref{eq25}) provides a useful formula 
\begin{equation}
F=3\left(\frac{f}{  1 {\rm kHz}}\right)^2\left( \frac{h}{
10^{-22}}\right)^2 \frac{{\rm ergs}}{ {\rm cm}^2 \rm{sec}},
\label{eq26}
\end{equation}
 which can be used to estimate the flux on  Earth, given the
amplitude of the waves (on  Earth) and the frequency of the waves.

\subsection{Generation of gravitational waves}

  As  early  as  1918,  Einstein derived  the  quadrupole
formula  for  gravitational radiation. This  formula  states that
the wave amplitude $h_{ij}$ is proportional to the second time
derivative of the quadrupole moment of the source:
\begin{equation}
h_{ij}=\frac{2}{r} \frac{G}{c^4}{\ddot Q}_{ij}^{TT}\left(t-\frac{r}{c}\right) \label{eq27}
\end{equation}
 where
\begin{equation}
Q_{ij}^{TT}(x)=\int \rho\left(x^i x^j- \frac{1}{3}\delta^{ij}r^2\right)d^3 x \label{eq28}
\end{equation}
is  the quadrupole moment in the TT gauge, evaluated at  the
retarded  time $t-r/c$  and $\rho$  is the matter  density  in  a
volume element $d^3 x$ at the position $x^i$. This result is quite
accurate for all sources,  as  long as the reduced wavelength
${\tilde \lambda}=\lambda/2\pi$ is much longer than the source
size $R$. It should be pointed out that the above result  can be
derived via a quite cumbersome calculation in which we solve the
wave equation (\ref{eq6}) with  a source term $T_{\mu\nu}$ on the
right-hand side. In the course of such a derivation, a number of
assumptions must be used. In particular, the observer must be
located at a distance $r \gg {\tilde \lambda} $, far greater  than
the reduced wavelength, in the so called  ``radiation zone'' and
$T_{\mu\nu}$ must not change very quickly.

     Using the formulae (\ref{eq24}) and (\ref{eq25}) for the energy carried
by  gravitational  waves, one can derive the  luminosity  in
gravitational  waves as a function of the  third-order  time
derivative  of  the quadrupole moment tensor.  This  is  the
quadrupole formula
\begin{equation}
L_{GW}=-\frac{dE}{dt} =\frac{1}{5}\frac{G}{c^5}\langle \frac{\partial^3
Q_{ij} }{ \partial t^3} \  \frac{\partial^3 Q_{ij}}{ \partial
t^3}\rangle. \label{eq29}
\end{equation}
Based  on  this formula, we derive some additional formulas, which
provide order of magnitude estimates for the amplitude of the
gravitational  waves and  the corresponding  power output of a
source. First, the quadrupole moment of a system is approximately
equal to the mass $M$ of the part of the system that moves, times
the square of the size $R$ of the system. This means that the
third-order time derivative of the quadrupole moment is
\begin{equation}
\frac{\partial^3 Q_{ij}}{ \partial t^3}\sim \frac{M R^2 }{ T^3} \sim
\frac{M v^2}{  T} \sim \frac{E_{\rm ns}}{ T}, \label{eq30}
\end{equation}
where $v$  is  the mean velocity of the moving parts,  $E_{\rm
ns}$ is the kinetic  energy  of the component of the source's
internal motion which is non spherical, and $T$ is the time scale
for a mass to move from one side of the system to the other.  The
time scale (or  period)  is actually  proportional  to the inverse
of the square root of the mean density of the system
\begin{equation}
T\sim \sqrt{R^3/GM} . \label{eq31}
\end{equation}
This relation provides   a rough   estimate   of   the
characteristic frequency of  the system $f=2\pi/T$. 
For  example,  for a non-radially oscillating  neutron  star with
a mass of roughly 1.4M$_\odot$ and a radius of  12km,  the frequency  of oscillation
which is directly related to the  frequency  of the  emitted
gravitational waves, will be roughly 2kHz. Similarly, for an oscillating black hole of the same mass we get a characteristic frequency of 10kHz.

 Then,   the luminosity  of gravitational waves of  a given
source is approximately
\begin{equation}
L_{GW}\approx \frac{G^4}{ c^5}\left(\frac{M}{R}\right)^5\sim
\frac{G}{c^5}\left(\frac{M}{R}\right)^2v^6\sim \frac{c^5}{G}\left(\frac{R_{\rm
Sch}}{R}\right)^2\left(\frac{v}{c}\right)^6 \label{eq32}
\end{equation}
and
\begin{equation}
h \approx \frac{1}{c^2}\left( \frac{GM}{r}\right) \left( \frac{R_{\rm Sch}}{R}\right) \approx \frac{2}{c^2}\left( \frac{GM}{r}\right)\left( \frac{v}{c}\right)^2
\end{equation}
where $R_{\rm Sch}=2GM/c^2$ is  the Schwarzschild radius of the
source. It is obvious that  the maximum  values of the amplitude and the  luminosity
of gravitational waves  can be  achieved  if the source's
dimensions are of the order of its Schwarzschild radius and the
typical velocities of the components of the system  are of  the
order of the speed of light. This explains why  we expect the best
gravitational wave  sources to be highly relativistic compact
objects. The above formula sets also an upper limit on the power
emitted by a source, which for $R\sim R_{\rm Sch}$  and $v\sim c$
is
\begin{equation}
L_{GW}\sim c^5/G= 3.6\times 10^{59} {\rm ergs/sec}.
\end{equation}
This is an immense power, often called the {\em luminosity of the
universe}.

 Using  the above order-of-magnitude estimates,  we  can
get a rough estimate of the amplitude of gravitational waves at a
distance $r$ from the source:
\begin{equation}
h\sim \frac{G}{c^4}\frac{E_{\rm ns}}{ r} \sim \frac{G}{
c^4}\frac{\varepsilon \, E_{\rm kin}}{ r} \label{eq34}
\end{equation}
where $\varepsilon E_{\rm kin}$ (with $0\le \varepsilon \le 1$),
is the fraction of kinetic energy of the source that  is  able to
produce gravitational waves.  The factor $\varepsilon$ is  a
measure of the asymmetry of the source and implies that only a
time varying quadrupole moment will emit gravitational waves. For
example, even if a huge amount of kinetic energy  is involved in a
given explosion and/or implosion,  if the  event takes  place in a
spherically symmetric manner, there will be no gravitational
radiation.

     Another  formula  for  the amplitude  of  gravitational
waves  relation can be derived from the flux  formula
(\ref{eq26}). If, for example, we consider an event (perhaps a
supernovae explosion) at  the Virgo cluster during  which  the
energy equivalent  of $10^{-4} M_\odot$ is released in
gravitational waves at a frequency of 1~kHz, and with signal
duration  of the order of 1~msec, the amplitude of the
gravitational waves on  Earth will be
\begin{equation}
h\approx 10^{-22}\left(\frac{E_{GW}}{  10^{-4} M_\odot c^2
}\right)^{1/2} \left(\frac{f}{1 {\rm kHz}}\right)^{-1} \left(\frac{\tau
}{1 {\rm msec}}\right)^{-1/2} \left(\frac{r}{15 {\rm
Mpc}}\right)^{-1}. \label{eq35}
\end{equation}
For a detector with arm length of 4~km we are looking for changes
in the arm length of the order of
$$
\Delta \ell =h \cdot \ell = 10^{-22}\cdot 4{\rm km} = 4\times
10^{-17} {\rm cm} \, !
$$
 This small number explains why all detection efforts till today were
not successful.

If the signal analysis is based on matched filtering, the \emph{effective amplitude} 
improves roughly as the square root of the number of observed cycles $n$. 
Using $n \approx f \tau $ we get 
\be
h_{\rm c} 
\approx  10^{-22} \left( \frac{ E_{GW}}{ 10^{-3} M_\odot c^2} \right)^{ 1/2}
\left(\frac{ f}{ 1 \mbox{ kHz} }\right)^{-1/2} 
\left( \frac{ r}{15 \mbox{ Mpc}}\right)^{-1}
\label{heff}
\ee
 We see that 
the ``detector sensitivity'' essentially depends only on the 
radiated energy, the characteristic frequency and the distance to the source.
That is, in order to obtain a rough estimate of the relevance of a 
given gravitational-wave source at a given distance we only need to estimate the
frequency and the radiated energy. Alternatively, if we know the energy released 
we can work out the distance at which these sources can be detected.  

\subsection{Gravitational wave detection}

One often classifies
gravitational-wave sources by the nature of the waves. This is convenient 
because the different classes require different approaches to the data-analysis
problem;

\begin{itemize}

\item {\em Chirps}. As a binary system radiates gravitational waves and 
loses energy the two constituents spiral closer together. As the separation
decreases the gravitational-wave amplitude increases, leading to a 
characteristic ``chirp'' signal. 

\item {\em Bursts}. Many scenarios lead to burst-like gravitational waves.
A typical example would be black-hole oscillations excited during binary 
merger. 

\item {\em Periodic}. Systems where the gravitational-wave backreaction
leads to a slow evolution (compared to the observation time)
may radiate persistent waves with a virtually constant frequency.
This would be the gravitational-wave analogue of the radio 
pulsars. 

\item {\em Stochastic}. A stochastic (non-thermal) 
background of gravitational waves
is expected to have been generated following the Big Bang. 
 One may also have to deal with 
stochastic gravitational-wave signals when the sources are too 
abundant for us to distinguish them as individuals.

\end{itemize}

Given that the weak signals are going to be buried in detector noise, 
we need to obtain as accurate theoretical models as possible.
The rough order of magnitude estimates we just derived will certainly 
not be sufficient, even though they provide an indication 
as to whether it is worth spending the time and effort 
required to build a detailed model. Such source models
are typically obtained using either
\begin{itemize}
\item Approximate perturbation techniques, eg. expansions in 
small perturbations away from a known solution to the Einstein equations, the 
archetypal case being black-hole and neutron star oscillations.

\item Post-Newtonian approximations, essentially an expansion 
in the ratio between a characteristic velocity of the system and the 
speed of light, 
most often used to model
the inspiral phase of a compact binary system.

\item Numerical relativity, where the Einstein equations are formulated
as an initial-value problem and solved on the computer. This is the only 
way to make progress in situations where the full nonlinearities of the theory 
must be included, eg. in the merger of black holes and neutron stars
or a supernova core collapse. 

\end{itemize} 

 The  first  attempt to detect gravitational  waves  was
undertaken  by  the pioneer Joseph Weber  during  the  early
1960s. He  developed the first resonant  mass  (bar) detector  and
inspired many  other  physicists  to  build  new detectors and to
explore  from  a  theoretical  viewpoint possible cosmic sources
of gravitational radiation.   When  a
gravitational wave hits such a device, it causes the bar  to
vibrate.  By  monitoring this vibration, we can  reconstruct the
true  waveform. 
The next step, was to monitor the change of the distance between 
two fixed points by  a passing-by
gravitational  wave. This can be done by  using
laser interferometry. The  use  of interferometry  is probably
the most decisive  step  in  our attempt  to  detect
gravitational  wave  signals.  
 Although the basic principle of such detectors is  very
simple,  the sensitivity of detectors is limited by  various
sources of noise. The internal noise of the detectors can be
Gaussian  or non-Gaussian. The non-Gaussian noise may  occur
several  times  per  day  such as  strain  releases  in  the
suspension  systems  which isolate the  detector  from  any
environmental mechanical source of noise, and the  only  way to
remove this type of noise is via comparisons of the data streams
from various detectors. The so-called Gaussian noise obeys  the
probability distribution of Gaussian  statistics and  can  be
characterized  by a  {\em spectral  density} $S_n(f)$.  The
observed signal at the output of a detector consists of  the true
gravitational wave strain $h$  and  Gaussian noise.  The optimal
method to detect a gravitational wave signal  leads to the
following signal-to-noise ratio:
\begin{equation}
\left(\frac{S}{N} \right)^2_{\rm opt} = 2 \int_0^\infty \frac{|{\tilde
h}(f)|^2}{S_n(f)}df, \label{eq42}
\end{equation}
where ${\tilde h}(f)$ is the Fourier transform of the signal
waveform.  It is clear from this expression that  the  sensitivity
of gravitational wave detectors is limited by noise.

 In  reality, the efficiency of a resonant bar  detector
depends  on several other parameters. Here, we will  discuss only
the  more fundamental ones. Assuming perfect isolation of  the
resonant bar detector from any external  source  of noise
(acoustical, seismic, electromagnetic),  the  thermal noise  is
the  only factor limiting our ability  to  detect gravitational
waves. Thus, in order to detect a signal,  the energy  deposited
by the gravitational wave  every $\tau$  seconds should   be
larger than  the  energy  $k T$ due   to   thermal fluctuations.
This leads to a  formula  for  the  minimum detectable  energy
flux of gravitational  waves,   which, following  equation
(\ref{eq25}), leads into a  minimum  detectable strain amplitude
\begin{equation}
h_{\rm min} \le \frac{1}{ \omega_0 L Q}\sqrt{\frac{15 k T}{  M}}
\label{eq46}
\end{equation}
where $L$ and $M$ are the size and the mass of the resonant bar correspondingly and $Q$ is the quality factor of the material. 
During the last 20 years,  a  number  of  resonant   bar
detectors  have  been  in  nearly  continuous  operation  in
several   places  around  the  world.  They  have   achieved
sensitivities of a few times $10^{-21}$, but still there has been
no clear evidence of gravitational wave detection. 

   A  laser interferometer is an alternative gravitational
wave  detector  that  offers the possibility  of  very  high
sensitivities  over a broad frequency band. Originally,  the idea
was to construct a new type of resonant detector  with much larger
dimensions. 
 Gravitational waves that
are  propagating perpendicular  to  the  plane  of  the
interferometer  will increase the length of one arm of the
interferometer, and at the  same  time will shorten the other arm,
and vice  versa. This technique of monitoring the waves is based
on Michelson interferometry. L-shaped interferometers are
particularly suited to the detection of gravitational waves due to
their quadrupolar nature. For extensive reviews refer to \cite{Saulson, Blair, HR2000,MM2008}.

The US project named LIGO
\cite{Ligo}(Laser Interferometer Gravitational Wave Observatory) consists
of two detectors with arm length of 4~Km, one in Hanford,
Washington, and one  in Livingston, Louisiana. The detector in
Hanford includes, in the same vacuum system, a second detector with
arm length of 2~km. The detectors are already in operation and they
achieved the designed sensitivity.
     The Italian/French EGO (VIRGO) detector \cite{Virgo} of arm-length 3~km at
Cascina near Pisa is designed to have better sensitivity  at lower
frequencies.  GEO600 is
the German/British detector built in Hannover \cite{Geo}. 
 The TAMA300\cite{Tama} detector in Tokyo has arm
lengths of 300~m and it was the first major interferometric
detector in operation. There are already plans for improving the
sensitivities of all the above detectors and the construction of
new interferometers in the near future.

Up to now LIGO has completed four science runs, S1 from  August - September 2002, S2 from  February - April 2003, S3 from October 2003 - January 2004, and S4 February - March 2005. These short science runs where interrupted with improvements which led LIGO to operate now with its designed sensitivity. The fifth science run, S5, (November 2005 - September 2007) surveyed a considerable larger volume of the Universe and set upper limits to a number of gravitational wave sources \cite{LIGO_background,LIGO_1, LIGO_2,LIGO_pulsar1,LIGO_pulsar2}. The ``enchanced'' LIGO interferometer is expected to commence an S6 science run in 2009 and will survey a volume of space eight times as great as the current LIGO. In 2011, the LIGO interferometers will be shut down for decommisioning in order to install advanced interferometers. With these advanced interferometers LIGO is expected to operate with ten times the current sensitivity which means a factor of 1000 increase in the volume of the Universe surveyed by 2014.

\section{Sources of gravitational waves}
\subsection{Radiation from binary systems}

 Among the  most interesting sources of gravitational
waves  are  binaries.  The  inspiralling  of  such  systems,
consisting  of black holes or neutron stars, is, as  we  will
discuss   later,   the  most  promising   source   for   the
gravitational wave detectors. Binary systems  are  also  the
sources  of gravitational waves whose dynamics we understand the
best. They  emit  copious  amounts   of gravitational
radiation,  and  for  a given  system we know quite accurately
the amplitude and  frequency of the gravitational waves in
terms of the masses of the two bodies and their separation. 

The gravitational-wave signal from inspiraling binaries is 
approximatelly sinusoidal, with a frequency 
which is twice the orbital frequency of the binary.  According  to equation (\ref{eq29}) the gravitational radiation luminosity of the system is
\begin{equation}
L^{\rm GW}=\frac{32}{5} \frac{G}{c^5}\mu^2 a^4 \Omega^6 = \frac{32}{
5} \frac{G^4}{c^5} \frac{M^3 \mu^2}{ a^5}, \label{eq38}
\end{equation}
where $\Omega$ is the orbital angular velocity, $a$ is the distance between the two bodies, $\mu=M_1 M_2/M$ is
the reduced mass of the system and $M=M_1+M_2$ is its total mass.
In order to obtain the last part of the relation,  we have
used Kepler's third law, $\Omega^2=GM/a^3$. As the gravitating
system loses energy by emitting radiation, the distance between
the two bodies shrinks and the orbital frequency increases accordingly (${\dot T}/T=1.5
{\dot a}/a$). Finally, the amplitude of the gravitational waves is
\begin{equation}
h=5\times 10^{-22} \left(\frac{M}{2.8 M_\odot}\right)^{2/3}
\left(\frac{\mu}{0.7M_\odot}\right) \left(\frac{f}{100 {\rm
Hz}}\right)^{2/3} \left(\frac{15 {\rm Mpc}}{ r}\right). \label{eq41}
\end{equation}
In all these formulae we have assumed that the orbits are
circular.

As the binary system 
evolves the orbit shrinks and the frequency increases in the characteristic chirp. 
Eventually, depending on the masses of the binaries, the frequency of the 
emitted gravitational waves will enter the bandwidth of the detector 
at the  low-frequency end and will evolve quite fast towards higher frequencies. 
A system consisting  of  two  neutron  stars
will  be detectable by LIGO when the frequency of the
gravitational waves is  $\sim$10Hz until the final coalescence
around 1 kHz. This process will last for about 15 min and the
total number of observed cycles will be of the order of $10^4$,
which leads to  an enhancement of the detectability by  a  factor
100 (remember $h_c \sim \sqrt{n}h$). 
Binary neutron  star systems and  binary  black  hole systems
with masses of the order of 50M$_\odot$ are  the primary sources
for LIGO. Given the anticipated sensitivity of LIGO, binary black
hole systems are the most  promising sources and could be detected
as far as 200 Mpc  away.  For the present
estimated sensitivity of LIGO the event rate 
is probably a few per year,
but future improvements of detector sensitivity (the LIGO II
phase) could lead  to  the detection  of  at  least one event per
month.  Supermassive black hole  systems of a few million solar
masses  are  the primary source for LISA. These binary systems are
rare,  but due  to  the huge amount of energy released, they
should  be detectable  from as far away as the boundaries of the
observable universe. Finally, the recent discovery of the highly relativistic 
binary pulsar J0737-3039 \cite{Burgay03} enchanced considerably 
the expected coalescence event rate of NS-NS binaries \cite{Kalogera04}. 
The event rate for initial LIGO is in the best case 
0.2 per year while advanced LIGO might be able to detect 20-1000 events per year.


\begin{figure}
\vskip 0.4cm
 \centerline{\includegraphics[height=6.5cm]{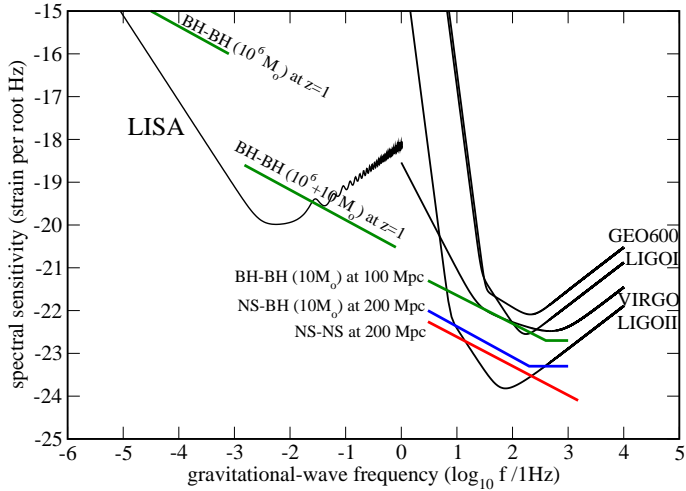}}
\medskip
\caption{Estimated signal strengths for various inspiralling binaries relevant for 
ground- and space-based detectors. } 
\label{bins}\end{figure}

Depending on the high-density EOS and their initial masses, the
outcome of the merger of two neutron stars may not always be a black
hole, but a hypermassive, differentially rotating compact star (even
if it is only temporarily supported against collapse by differential
rotation). A recent detailed simulation\cite{Shibata05b} in full GR
has shown that the hypermassive object created in a binary NS merger
is nonaxisymmetric. The nonaxisymmetry lasts for a large number of
rotational periods, leading to the emission of gravitational waves
with a frequency of $~3$kHz and an effective amplitude of $\sim
6-7\times 10^{-21}$ at a large distance of 50Mpc. Such large
effective amplitude may be detectable even by LIGO II at this high
frequency.

The tidal disruption of a NS by a BH \cite{Vallisneri} or the merging
of two NSs \cite{Rasio} may give valuable information for the radius
and the EoS if we can recover the signal at frequencies higher than 1
kHz.

\subsection{Gravitational collapse}

 One of the most spectacular astrophysical events
is the core collapse of massive stars, leading to the
formation of a neutron star (NS) or a black hole (BH).
The outcome of core collapse depends sensitively on several
factors: mass, angular momentum and metallicity of progenitor,
existence of a binary companion, high-density equation of
state, neutrino emission, magnetic fields, etc. Partial
understanding of each of the above factors is
emerging, but a complete and consistent theory for core
collapse is still years away.

Roughly speaking, isolated stars more massive than $\sim
8-10M_\odot$ end in core collapse and  $\sim 90\%$ of them are
stars with masses $\sim
8-20M_\odot$. After core bounce, most of the material is ejected and
if the progenitor star has a mass $M\lesssim 20M_\odot$ a
neutron star is left behind. On the other hand, if $M\gtrsim
20M_\odot$ fall-back accretion increases the mass of the
formed proton-neutron star (PNS), pushing it above the maximum
mass limit, which results in the formation of a black hole.
Furthermore, if the progenitor star has a mass of roughly $M\gtrsim
45M_\odot$, no supernova explosion is launched and the star collapses
directly to a BH\cite{Fryer99}.

The above picture is, of course, greatly simplified. In reality,
the metallicity of the progenitor, the angular momentum of the
pre-collapse core and the presence of a binary companion will
decisively influence the outcome of core collapse\cite{Fryer2003}.
Rotation influences the collapse by changing
dramatically the properties of the convective region above the
proto-neutron star core. Centrifugal forces slow down infalling
material in the equatorial region compared to materiall falling in
along the polar axis, yielding a weaker bounce. This
asymmetry between equator and poles also strongly influences the
neutrino emmission and the revival of the stalled shock by neutrinos
\cite{Fryer2004,Burrows2005}.

The supernova event rate is 1-2 per century per galaxy
\cite{Cappellaro} and about 5-40\% of them produce BHs in
delayed collapse (through fall-back accretion), or direct
collapse\cite{FryerKalogera}.
 
 Of considerable importance
is the {\em initial rotation rate} of proto-neutron stars, since (as
will be detailed in the next sections) most mechanisms for
emission of detectable gravitational waves from compact
objects require very rapid rotation at birth (rotational
periods of the order of a few milliseconds or less).
Since most massive stars have non-negligible rotation rates
(some even rotate near their break-up limit), simple conservation
of angular momentum would suggest a proto-neutron star
to be strongly differentially rotating with very high rotation
rates and this picture is supported by numerical simulations
of rotating core collapse\cite{FryerHeger,DFM2002b}.

Other ways to form a rapidly rotating proto-neutron star would
be through {\it fall-back accretion}\cite{Watts02},
through the {\it accretion-induced collapse of a white
dwarf} \cite{Hillebrandt1984,Liu2001,Fryer2002,Yoon05} or
through the merger of binary white dwarfs in globular
clusters\cite{Middleditch04}. It is also relevant to take into account
current gamma-ray-burst models. The {\it collapsar} \cite{Woosley93}
model requires high rotation rates of a proto-black
hole \cite{Petrovic05}. In addition, a possible formation scenario for
magnetars involves a rapidly rotating PNS formed through
the collape of a very massive progenitor and some observational
evidence is already emerging \cite{Gaensler05}.

 Gravitational waves
from core collapse have a rich spectrum, reflecting the
various stages of this event. The initial signal is emitted due
to the {\it changing axisymmetric quadrupole moment} during collapse.
In the case of neutron star formation, the quadrupole moment
typically becomes larger, as the core spins up during contraction.
In contrast, when a rapidly rotating neutron star collapses to
form a Kerr black hole, the axisymmetric quadrupole moment
first increases but is finally reduced by a large factor when the
black hole is formed.

A second part of the gravitational wave signal is produced
when gravitational collapse is halted by the stiffening
of the equation of state above nuclear densities and the core bounces,
driving an outwards moving shock. The dense fluid undergoes motions
with relativistic speeds ($v/c\sim 0.2-0.4$) and a rapidly rotating
proto-neutron star thus oscillates in several of its
axisymmetric {\it normal modes of oscillation}.
This quasi-periodic part of the signal could last for hundreds of
oscillation periods, before being effectively damped.
If, instead, a black hole is directly formed, then black hole
quasinormal modes are excited, lasting for only a few
oscillation periods. A combination of neutron star and black
hole oscillations will appear if the proton-neutron star is not stable
but collapses to a black hole.

In a rotating PNS, nonaxisymmetric processes can
yield additional types of gravitational wave signals. Such processes
are {\it dynamical instabilities}, {\it secular
gravitational-wave driven instabilities} or {\it convection} inside
the PNS and in its surrounding hot envelope.
{\it Anisotropic neutrino emission}  is accompanied by a gravitational
wave signal. {\it Nonaxisymmetries} could
already be present in the pre-collapse core and become amplified
during collapse\cite{Fryer2004b}. Furthermore, if there
is persistent fall-back accretion onto a PNS or black hole, these can be brought into {\it ringing}.

Below, we discuss in more detail those processes which result in
high frequency gravitational radiation.

\subsubsection{Neutron star formation}
Core collapse as a potential source of GWs has been studied for more
than three decades (some of the most recent simulations can be
found in
\cite{Zwerger1997,Rampp1998,Fryer2002,DFM2002b,Ott2003,Kotake2003,Mueller2004,Shibata04,Shibata05,Sekiguchi05}). The main differences between the various
studies are the
progenitor models (slowly or rapidly rotating), equation of
state (polytropic or realistic), gravity (Newtonian
or relativistic) and neutrino emission (simple, sophisticated or
no treatment). In general, the gravitational wave signal from
neutron star formation is divided into a core bounce signal, a
signal due to convective motions and a signal due to anisotropic
neutrino emission.

 The core bounce signal is
produced due to rotational flattening and excitation of normal modes
of oscillations, the main contributions coming from the axisymmetric
quadrupole ($l=2$) and quasi-radial ($l=0$) modes (the latter
radiating through its rotationally acquired $l=2$ piece). If
detected, such signals will be a unique probe for the high-density
EOS of neutron stars \cite{AK1996,AK1998}. The strength of this
signal is sensitive to the available angular momentum in the
progenitor core. If the progenitor core is rapidly rotating, then
core bounce signals from Galactic supernovae ($d\sim 10$kpc) are
detectable even with the initial LIGO/Virgo sensitivity at
frequencies $\lesssim$1kHz. In the best-case scenario, advanced LIGO
could detect signals from distances of 1Mpc, but not from the Virgo
cluster ($\sim$15Mpc), where the event rate would be high. The
typical GW amplitude from 2D numerical simulations
\cite{DFM2002b,Ott2003,Burrows2007, Dimmelmeier2007, Ott2007,Ott2007a} for an observer located in the equatorial
plane of the source is \cite{Shibata04}
\begin{equation}
h\approx 9 \times 10^{-21}\varepsilon \left(\frac{ 10 {\rm
kpc}}{d}\right) \, ,
\end{equation}
where $\varepsilon \sim 1$ is the normalized GW amplitude. For
such rapidly rotating initial models, the total energy radiated
in GWs during the collapse is $\lesssim
10^{-6}-10^{-8} M_\odot c^2$.
If, on the other hand, progenitor cores are slowly rotating (due to
e.g. magnetic torques\cite{Spruit2002}), then the signal strength is
significantly reduced, but, in the best case, is still within reach of
advanced LIGO for galactic sources.

Normal mode oscillations, if excited in an equilibrium star at a
small to moderate amplitude, would last for hundreds to thousands of
oscillation periods, being damped only slowly by gravitational wave
emission or viscosity. However, the PNS immediately
after core bounce has a very different structure than a cold
equilibrium star. It has a high internal temperature and is
surrounded by an extended, hot envelope. Nonlinear oscillations
excited in the core after bounce can penetrate into the hot
envelope. Through this damping mechanism, the normal mode
oscillations are damped on a much shorter timescale (on the order of
ten oscillation periods), which is typically seen in the core
collapse simulations mentioned above.

 {\it Convection signal.}
The post-shock region surrounding a PNS is
convectively unstable to both low-mode and high-mode
convection. Neutrino emission also drives convection in this
region. The most realistic 2D simulations of core collapse
to date \cite{Mueller2004} have shown that the gravitational wave
signal from convection significantly exceeds the core bounce
signal for slowly rotating progenitors, being detectable with
advanced LIGO for galactic sources, and is detectable even
for nonrotating collapse. For slowly rotating collapse, there
is a detectable part of the signal in the high-frequency
range of 700Hz-1kHz, originating from convective motions that
dominate around 200ms after core bounce. Thus, if both a core
bounce signal and a convection signal would be detected in
the same frequency range, these would be well separated in time.

 {\it Neutrino signal.}
In many simulations the gravitational wave signature of
anisotropic neutrino emission has also been
considered \cite{Epstein78,Burrows96,Mueller97}. This type
of signal can be detectable by advanced LIGO
for galactic sources, but the main contribution is at
low frequencies for a slowly rotating progenitor \cite{Mueller2004}.
 For rapidly rotating progenitors, stronger contributions at
high frequencies could be present, but would probably be
burried within the high-frequency convection signal.

Numerical simulations of neutron star formation have gone
a long way, but a fully consistent 3D simulation including
relativistic gravity, neutrino emission and magnetic fields
is still missing. The combined treatment of these effects might not
change the above estimations by orders of magnitude but it will
provide more conclusive answers. There
are also issues that need to be understood such as pulsar kicks
(velocities exceeding 1000 km/s) which suggest that in a
fraction of newly-born NSs (and probably BHs) the formation
process may be strongly asymmetric \cite{Hoef02}.
Better treatment of the microphysics and construction of accurate
progenitor models for the angular momentum distributions are needed.
All these issues are under investigation by many groups.

\subsubsection{Black hole formation}

The gravitational-wave emission from the formation of a Kerr BH is
a sum of two signals: the {\it collapse signal} and the {\it BH
ringing}. The collapse signal is produced due to the changing
multipole moments of the spacetime during the transition from a
rotating iron core or PNS to a Kerr BH. A uniformly
rotating neutron star has an axisymmetric quadrupole moment given
by\cite{Laarakkers99}
\begin{equation}
Q=-a\frac{J^2}{M} \, ,
\end{equation}
where $a$ depends on the equation of state and is in the range
of $2-8$ for 1.4$M_\odot$ models. This is several times larger in
magnitude than the corresponding quadrupole moment of
a Kerr black hole ($a=1$). Thus, the {\it reduction} of the
axisymmetric
quadrupole moment is the main source of the collapse signal. Once
the BH is formed, it continues to oscillate in its axisymmetric
$l=2$ quasinormal mode (QNM), until all oscillation energy is radiated away and the
stationary Kerr limit is approached.

The numerical study of rotating collapse to BHs was pioneered by
Nakamura\cite{Nakamura81}
 but first waveforms and gravitational-wave estimates were obtained
by Stark and Piran\cite{Stark87}
. These simulations we performed in 2D, using approximate initial data
(essentially a spherical star to which angular momentum was
artificially added). A new 3D computation of the gravitational wave
emission from the collapse of unstable uniformly rotating
relativistic polytropes to Kerr BHs\cite{Baiotti05}
finds that the energy emitted is
\begin{equation}
\Delta E \sim 1.5 \times 10^{-6} (M/M_\odot),
\end{equation}
significantly less than the result of Stark and Piran.
Still, the collapse of an unstable 2$M_\odot$ rapidly rotating
neutron star leads to a characteristic gravitational-wave amplitude
$h_c \sim 3\times 10^{-21}$, at a frequency of $\sim 5.5$kHz, for an
event at 10kpc. Emission is mainly through the "+" polarization, with
the "$\times$" polarization being an order of magnitude weaker.

Whether a BH forms promptly after collapse or a delayed collapse
takes place depends sensitively on a number of factors, such as the
progenitor mass and angular momentum and the high-density EOS. The
most detailed investigation of the influence of these factors on
the outcome of collapse has been presented recently
in\cite{Sekiguchi05}, where it was found that shock formation
increases the threshold for
black hole formation by $\sim 20-40\%$, while rotation results in an
increase of at most 25\%.

\subsubsection{Black hole ringing through fall-back}

A black hole can form after core collapse if fall-back
accretion increases the mass of the PNS above the
maximum mass allowed by axisymmetric stability. Material falling back
after the black hole is formed excites the black hole QNMs of oscillation. If, on the other hand, the black hole is formed
directly through core collapse (without a core bounce taking place)
then most of the material of the progenitor star is accreted at
very high rates ($\sim 1-2M_\odot$/s) into the hole.
In such {\it hyper-accretion} the black hole's QNMs can be excited for as long as the process lasts and
until the black hole becomes stationary. Typical frequencies of the
emitted GWs are in the range 1-3kHz for $\sim 3-10 M_\odot$ BHs.

The frequency and the damping time of the oscillations for the
$l=m=2$
mode can be estimated via the relations \cite{Echeverria}
\begin{eqnarray}
\sigma&\approx& 3.2 {\rm kHz} \
M_{10}^{-1}\left[1-0.63(1-a/M)^{3/10}\right] \\
 Q&=&\pi \sigma
\tau \approx 2\left(1-a\right)^{-9/20} \label{bhqnm}
\end{eqnarray}
These relations together with similar ones either for the 2nd QNM or
the $l=2$, $m=0$ mode can uniquely determine the mass $M$ and
angular momentum parameter $a$ of the BH if the frequency and the
damping time of the signal have been accurately extracted
\cite{Finn,Nakano2003,Dryer}. The amplitude of the ring-down waves
depends on the BH's initial distortion, i.e. on the nonaxisymmetry
of the blobs or shells of matter falling into the BH. If matter of
mass $\mu$ falls into a BH of mass $M$, then the gravitational wave
energy is roughly
\begin{equation}
\Delta E \gtrsim
\varepsilon \mu c^2(\mu/M)
\end{equation}
 where $\varepsilon$ is related
to the degree of asymmetry and could be $\varepsilon
\gtrsim 0.01$ \cite{Davis1971}. This
leads to an effective GW amplitude
\begin{equation}
h_{\rm eff}\approx 2\times 10^{-21}\left(\frac{\varepsilon}{  0.01}
\right)\left(\frac{10 {\rm Mpc}}{ d}\right)\left( \frac{\mu}{
M_\odot}\right)
\end{equation}

\subsection{Rotational instabilities}

If proto-neutron stars rotate rapidly, nonaxisymmetric {\it dynamical
instabilities} can develop. These arise from non-axisymmetric
perturbations having angular dependence $e^{i
m\phi}$ and are of two different types: the {\it classical bar-mode}
instability and the more recently discovered {\it low-$T/|W|$
bar-mode} and {\it one-armed spiral} instabilities, which appear to be
associated to the presence of
corotation points. Another class of nonaxisymmetric instabilities are
{\it secular instabilities}, driven by dissipative effects, such as
fluid viscosity or gravitational radiation.

\subsubsection{Dynamical instabilities}

The classical $m=2$ bar-mode instability is excited in Newtonian
stars when the ratio $\beta=T/|W|$ of the rotational kinetic energy
$T$ to the gravitational binding energy $|W|$ is larger than
$\beta_{\rm dyn}=0.27$.
The instability grows on a dynamical time scale (the time
that a sound wave needs to travel across the star) which is about one
rotational period and may last from 1 to 100 rotations depending on
the degree of differential rotation in the PNS.

The bar-mode instability can be excited in a hot PNS, a few
milliseconds after core bounce, or alternatively, it could also be
excited a few tenths of seconds later,
when the PNS cools due to neutrino emission and contracts
further, with $\beta$ becoming larger than the threshold
$\beta_{dyn}$ ( $\beta$ increases roughly as $\sim 1/R$ during
contraction). The amplitude of the emitted
gravitational waves can be estimated as $h\sim M R^2 \Omega^2/d$,
where $M$ is the mass of the body, $R$ its size, $\Omega$ the
rotation rate and $d$ the distance of the source. This leads to an
estimation of the GW amplitude
\begin{equation}
h \approx 9 \times 10^{-23} \left(\frac{\epsilon}{0.2} \right)
\left(\frac{f}{3 {\rm kHz}}\right)^2 \left(\frac{15 {\rm
Mpc}}{d}\right) M_{1.4} R_{10}^2.
\end{equation}
where $\epsilon$ measures the ellipticity of the bar, $M_{1.4}$ is measured
in units of $1.4 M_\odot$ and $R$ is measured in units of 10km.
Notice that, in uniformly rotating Maclaurin spheroids,  the
GW frequency $f$ is twice the rotational frequency $\Omega$. Such a
signal is detectable only from sources in our galaxy or the nearby
ones (our Local Group). If the sensitivity of the detectors is
improved in the kHz region, signals from the Virgo cluster could
be detectable. If the bar persists for many ($\sim$ 10-100) rotation
periods, then even signals from distances considerably larger than
the Virgo cluster will be detectable. Due to the requirement of
rapid rotation, the event rate of the classical dynamical
instability is considerably lower than the SN event rate.

The above estimates rely on Newtonian calculations; GR enhances the
onset of the instability, $\beta_{\rm
dyn}\sim 0.24$ \cite{SBS2000,Saijo01} and somewhat lower than that
for large compactness (large $M/R$). Fully relativistic
dynamical simulations of this instability have been obtained,
including detailed waveforms of the associated gravitational wave
emission. A detailed investigation of the required initial conditions
of the progenitor core, which can lead to the onset of the dynamical
bar-mode instability in the formed PNS, was presented
in \cite{Shibata04}. The amplitude of gravitational waves  due to
the bar-mode
instability was found to be larger by
an order of magnitude, compared to the axisymmetric core
collapse signal.

\vskip 0.4cm

\noindent{\it Low-$T/|W|$ instabilities.}
The {\it bar-mode} instability may be excited for significantly
smaller $\beta$, if centrifugal forces produce a peak in the density
off the source's rotational center \cite{Centrella2001}. Rotating stars
with a high degree of differential rotation are also dynamically
unstable for significantly lower $\beta_{\rm dyn}\gtrsim 0.01$
\cite{Shibata2002,Shibata2003}. According to this scenario the
unstable neutron
star settles down to a non-axisymmetric quasi-stationary state which
is a strong emitter of quasi-periodic gravitational waves
\begin{equation}
h_{\rm eff} \approx 3\times 10^{-22} \left(\frac{R_{\rm eq}}{  30 {\rm
km}} \right) \left(\frac{f}{800 {\rm Hz}}\right)^{1/2} \left(\frac{100
{\rm Mpc} }{ d}\right) M_{1.4}^{1/2} .
\end{equation}
The bar-mode instability of differentially rotating neutron stars is
an excellent source of gravitational waves, provided the high degree
of differential rotation that is required can be realized. One
should also consider the effects of viscosity and magnetic fields.
If magnetic fields enforce uniform rotation on a short timescale,
this could have strong consequences regarding the appearance and
duration of the dynamical nonaxisymmetric instabilities.

An $m=1$ {\it one-armed spiral} instability has also been shown to
become unstable in PNS, provided that the
differential rotation is sufficiently strong
\cite{Centrella2001,SBM2003}. Although it is dominated by a
``dipole" mode, the instability has a spiral character, conserving
the center of mass. The onset of the instability appears to be
linked to the presence of corotation points \cite{Saijo05}
 (a similar link to corotation points has been proposed for
 the low-$T/|W|$ bar mode instability \cite{Watts04,PSK08}
) and requires a very high degree of differential rotation (with
matter on the axis rotating at least 10 times faster than
matter on the equator). The $m=1$ spiral instability was recently
observed in simulations of rotating core collapse, which started
with the core of an evolved 20$M_\odot$ progenitor star to which
differential rotation was added \cite{Ott05}. Growing from noise level
($\sim 10^{-6}$) on a timescale
of ~5ms, the $m=1$ mode reached its maximum amplitude after
$\sim 100$ms.
Gravitational waves were emitted through the excitation
of an $m=2$ nonlinear harmonic at a frequency of $\sim 800$Hz with
an amplitude comparable to the core-bounce axisymmetric signal.

\subsubsection{Secular gravitational-wave-driven instabilities}

In a nonrotating star, the forward and backward moving modes
of same $(l,|m|)$ (corresponding to $(l,+m)$ and $(l,-m)$) have
eigenfrequencies $\pm |\sigma|$. Rotation splits this degeneracy
by an amount $\delta \sigma \sim m \Omega$ and both the
prograde and retrograde modes are dragged forward by the stellar
rotation. If the star spins sufficiently rapidly, a
mode which is retrograde (in the frame rotating with the star)
will appear as prograde in the inertial frame (a nonrotating observer
at infinity). Thus, an inertial observer sees GWs with positive
angular momentum emitted by the retrograde mode, but since the
perturbed fluid rotates slower than it would in the absence of the
perturbation, the angular momentum of the mode in the rotating
frame is negative. The emission of
GWs consequently makes the angular momentum of the mode increasingly
negative, leading to the instability. A mode is unstable when
$\sigma(\sigma-m\Omega) < 0$.
This class of {\em frame-dragging instabilities} is usually referred
to as Chandrasekhar-Friedman-Schutz \cite{Chandra70,Friedman78} (CFS)
instabilities.

\vskip 0.4cm

\noindent{\it $f$-mode instability.}
In the Newtonian limit, the $l=m=2$ $f$-mode (which has the shortest
growth time of all polar fluid modes) becomes unstable when
$T/|W|>0.14$, which is near or even above the mass-shedding
limit for typical polytropic EOSs used to model uniformly rotating
neutron stars. Dissipative effects (e.g. shear and bulk viscosity or
mutual friction in
superfluids) \cite{Cutler87,Lindblom79,Ipser91,Lindblom95}
leave only a small instability window near mass-shedding, at
temperatures of $\sim 10^9$K. However,
relativistic effects strengthen the instability considerably, lowering
the required $\beta$ to  $\approx 0.06-0.08$ \cite{SF1998,MSB99}
for most realistic EOSs and masses of $\sim 1.4M_\odot$ (for higher
masses, such as hypermassive stars created in a binary NS merger,
the required rotation rates are even lower).

Since PNSs rotate differentially, the above limits derived under
the assumption of uniform rotation are too strict. Unless uniform
rotation is enforced on a short timescale, due to e.g. magnetic
braking \cite{Liu04},
the $f$-mode instability will develop in a differentially rotating
background, in which the required $T/|W|$ is only somewhat larger
than the corresponding value for uniform rotation \cite{Yoshida}, but
the mass-shedding limit is dramatically relaxed. Thus, in a
differentially
rotating PNS, the $f$-mode instability window is huge, compared to
the case of uniform rotation and the instability can develop provided
there is sufficient $T/|W|$ to begin with.

The $f$-mode instability is an excellent source of GWs.
Simulations of its nonlinear development in the ellipsoidal
approximation \cite{Lai95}
have shown that the mode can grow to a large nonlinear amplitude,
modifying the background star from an axisymmetric shape to a
differentially rotating ellipsoid. In this modified background the
$f$-mode amplitude saturates and the ellipsoid becomes a strong
emitter of gravitational waves, radiating away angular momentum
until the star is slowed-down towards a stationary state. In the
case of uniform density ellipsoids, this stationary state is the
Dedekind ellipsoid, i.e. a nonaxisymmetric ellipsoid with internal
flows but with a stationary (nonradiating) shape in the inertial
frame. In the ellipsoidal approximation, the nonaxisymmetric
pattern radiates gravitational waves sweeping through the LIGO II
sensitivity window (from 1kHz down to about 100Hz) which could
become detectable out to a distance of more than 100Mpc.

Two recent hydrodynamical simulations \cite{ShibataKarino04,Ou04}
 (in the Newtonian limit
and using a post-Newtonian radiation-reaction potential)
essentially confirm this picture. In \cite{ShibataKarino04}
  a differentially rotating, $N=1$ polytropic model with a
 large $T/|W| \sim 0.2-0.26$ is chosen as the initial equilibrium
 state. The main difference
of this simulation compared to the ellipsoidal approximation
comes from the choice of EOS. For $N=1$ Newtonian polytropes
it is argued that the secular evolution cannot lead to 
a stationary Dedekind-like state. Instead,
the $f$-mode instability will continue to be
active until all nonaxisymmetries are radiated away and an
axisymmetric shape is reached. This conclusion should be
checked when relativistic effects are taken into account,
since, contrary to the Newtonian case, relativistic $N=1$
uniformly rotating polytropes {\it are} unstable to the $l=m=2$
$f$-mode \cite{SF1998}
 -- however it is not possible to date to
construct relativistic analogs of Dedekind ellipsoids.

In the other recent simulation \cite{Ou04}
, the initial state was chosen to be a uniformly rotating, $N=0.5$
polytropic model with $T/|W|\sim 0.18$. Again, the main conclusions
reached in {}\cite{Lai95}
 are confirmed, however, the assumption of uniform
initial rotation limits the available angular momentum that can be
radiated away, leading to a detectable signal only out to about
$\sim 40$Mpc. The star appears to be driven towards a Dedekind-like
state, but after about 10 dynamical periods the shape is disrupted
by growing short-wavelength motions, which are suggested to arise
because of a shearing type instability, such as the elliptic flow
instability \cite{Lifschitz93}.

{\it $r$-mode instability.} Rotation does not only shift
the spectra of polar modes; it also lifts the degeneracy of axial
modes, and gives rise to a new family of {\em inertial} modes, of which
the $l=m=2$ $r$-mode is a special member. The restoring force for
these oscillations is the Coriolis force. Inertial modes are
primarily velocity perturbations. The frequency of the $r$-mode in
the rotating frame of reference is $\sigma = 2 \Omega /3$. According
to the criterion for the onset of the CFS instability, the $r$-mode
is unstable for any rotation rate of the
star \cite{Andersson98,FriedmanMorsink}. For temperatures between
$10^{7}-10^{9}$K and rotation rates larger than 5-10\% of the Kepler
limit, the growth time of the unstable mode is smaller than the
damping times of the bulk and shear viscosity \cite{LOM98,AKS99}. The
existence of a solid crust  or of hyperons in the core \cite{LO2002}
 and magnetic fields \cite{Rezzolla1,Rezzolla2}, can also significantly affect the
onset of the instability (for extended reviews see
\cite{AK2001,Nils2003}). The suppression of the $r$-mode instability
by the presence of hyperons in the core is not expected to operate
efficiently in rapidly rotating stars, since the central density is
probably too low to allow for hyperon formation. Moreover, a recent
calculation\cite{vanDalen03} finds the contribution of hyperons to
the bulk viscosity to be two orders of magnitude smaller than
previously estimated. If accreting neutron stars in Low Mass X-Ray
Binaries (LMXB, considered to be the progenitors of millisecond
pulsars) are shown to reach high masses of $\sim 1.8M_\odot$, then
the EOS could be too stiff to allow for hyperons in the core (for
recent observations that support a high mass for some millisecond
pulsars see \cite{Nice03}).

The unstable $r$-mode grows exponentially until it saturates due to
nonlinear effects at some maximum amplitude $\alpha_{max}$. The
first computation of nonlinear mode couplings using second-order
perturbation theory suggested that the $r$-mode is limited to very
small amplitudes (of order $10^{-3}-10^{-4}$) due to transfer of
energy to a large number of other inertial modes, in the form of a
cascade, leading to an equilibrium distribution of mode
amplitudes \cite{Arras2002}. The small saturation values for the
amplitude are supported by recent nonlinear estimations
\cite{Sa2004,SaTome2005} based on the drift, induced by the r-modes,
causing differential rotation.  On the other hand, hydrodynamical
simulations of limited resolution showed that an initially
large-amplitude $r$-mode does not decay appreciably over several
dynamical timescales \cite{StergioulasFont}.  However, on a somewhat
longer timescale a catastrophic decay was observed \cite{Gressman02}
 indicating a transfer of energy to other modes, due to nonlinear
 mode couplings and suggesting that a hydrodynamical instability may
 be operating. A specific resonant 3-mode coupling was
identified \cite{Lin04} as
the cause of the instability and a perturbative analysis of the decay
rate suggests a maximum saturation amplitude $\alpha_{max} < 10^{-2}$.
A new computation using second-order perturbation
 theory finds that the catastrophic decay seen in the hydrodynamical
 simulations \cite{Gressman02,Lin04}
 can indeed be explained by a parametric instability operating
 in 3-mode couplings between the $r$-mode and two other inertial
 modes \cite{Brink04a,Brink04b,Brink05,Bondarescu2007}. Whether the maximum saturation
amplitude is set by a network
 of 3-mode couplings or a cascade is reached, is, however, still
 unclear.

A neutron star spinning down due to the $r$-mode instability
will emit gravitational waves of amplitude
\begin{equation}
h(t)\approx 10^{-21} \alpha \left( \frac{\Omega}{  1 {\rm
kHz}}\right)\left(\frac{100 {\rm kpc}}{ d}\right)
\end{equation}
Since $\alpha$ is small, even with LIGO II the signal is
undetectable at large distances (VIRGO cluster) where the SN
event rate is appreciable, but could be detectable after
long-time integration from a galactic event. However, if
the compact object is a strange star, then the instability may
not reach high amplitudes ($\alpha \sim 10^{-3}-10^{-4}$) but it
will persist for a few hundred years (due to the different
temperature dependence of viscosity in strange quark matter)
and in this case there might be
up to ten unstable stars in our galaxy at any time \cite{AJK2002}.
Integrating data for a few weeks could lead to an effective amplitude
$h_{\rm eff}\sim 10^{-21}$ for galactic signals at frequencies $\sim
700-1000$Hz. The frequency of the signal changes only slightly on a
timescale of a few months, so that the radiation is practically
monochromatic.

{\bf Other unstable modes.}  The CFS instability can also
operate for core g-mode oscillations \cite{Lai99}
 but also for {\em w}-mode oscillations, which are basically spacetime
 modes \cite{KRA2004}. In addition, the CFS instability can operate
 through other dissipative effects. Instead of the gravitational
radiation, any radiative mechanism (such as electromagnetic
radiation) can in principle lead to an instability.

\subsection{Accreting neutron stars in LMXBs}

Spinning neutron stars with even tiny deformations are interesting
sources of gravitational waves. The deformations might results from
various factors but it seems that the most interesting cases are the
ones in which the deformations are caused by accreting material. A
class of objects called Low-Mass X-Ray Binaries (LMXB) consist of a
fast rotating neutron star (spin $\approx 270-650$Hz) torqued by
accreting material from a companion star which has filled up its
Roche lobe. The material adds both mass and angular momentum to the
star, which, on timescales of the order of tenths of Megayears
could, in principle, spin up the neutron star to its break up limit.
One viable scenario \cite{Bildsten1998} suggests that the accreted
material (mainly hydrogen and helium) after an initial phase of
thermonuclear burning undergoes a non-uniform crystallization,
forming a crust at densities $\sim 10^8-10^9$g/cm$^3$. The
quadrupole moment of the deformed crust is the source of the emitted
gravitational radiation which slows-down the star, or halts the
spin-up by accretion.

An alternative scenario has been proposed by
Wagoner\cite{Wagoner1984} as a follow up of an earlier idea by
Papaloizou-Pringle \cite{PP1978}. The suggestion was that the spin-up
due to accretion might excite the $f$-mode instability before the
rotation reaches the breakup spin. The emission of gravitational
waves will torque down the star's spin at the same rate as the
accretion will torque it up, however, it is questionable whether the
$f$-mode instability will ever be excited for old, accreting neutron
stars. Following the discovery that the $r$-modes are unstable at
any rotation rate, this scenario has been revived independently by
Bildsten \cite{Bildsten1998} and Andersson, Kokkotas and Stergioulas
\cite{AKS1999}. The amplitude of the emitted gravitational waves
from such a process is quite small, even for high accretion rates,
but the sources are persistent and in our galactic neighborhood the
expected amplitude is
\begin{equation}
h\approx 10^{-27}\left(\frac{1.6\mbox{ms}}{P}
\right)^5\frac{1.5\mbox{kpc}}{D} \, .
\end{equation}
This signal is within reach of advanced LIGO with signal recycling
tuned at the appropriate frequency and integrating for a few
months\cite{ThorneCutler}. This picture is in practice more
complicated, since the growth rate of the $r$-modes (and
consequently the rate of gravitational wave emission) is a function
of the core temperature of the star. This leads to a thermal runaway
due to the heat released as viscous damping mechanisms counteract
the r-mode growth \cite{Levin99}. Thus, the system executes a limit
cycle, spinning up for several million years and spinning down in a
much shorter period. The duration of the unstable part of the cycle
depends critically on the saturation amplitude $\alpha_{max}$ of the
$r$-modes \cite{AJKS2000,Heyl}. Since current computations
\cite{Arras2002,SaTome2005} suggest an $\alpha_{max}\sim
10^{-3}-10^{-4}$, this leads to a quite long duration for the
unstable part of the cycle of the order of $\sim 1Myr$.

The instability window depends critically on the effect of the shear
and bulk viscosity and various alternative scenarios might be
considered. The existence of hyperons in the core of neutron stars
induces much stronger bulk viscosity, which suggests a much narrower
instability window for the $r$-modes and the bulk viscosity prevails
over the instability, even in temperatures as low as $10^8$K
\cite{LO2002}. A similar picture can be drawn if the star is
composed of ``deconfined" $u$, $d$ and $s$ quarks - a strange star
\cite{Madsen98}. In this case, there is a possibility that the
strange stars in LMXBs evolve into a quasi-steady state with nearly
constant rotation rate, temperature and mode amplitude
\cite{AJK2002} emitting gravitational waves for as long as the
accretion lasts. This result has also been found later for stars
with hyperon cores \cite{Wagoner,Reisenegger03}. It is interesting
that the stalling of the spin up in millisecond pulsars (MSPs) due
to $r$-modes is in good agreement with the minimum observed period
and the clustering of the frequencies of MSPs \cite{AJKS2000}.

\subsection{Gravitational-wave asteroseismology}

If various types of oscillation modes are excited during the
formation of a compact star and become detectable by gravitational
wave emission, one could try to identify observed frequencies with
frequencies obtained by mode-calculations for a wide parameter range
of masses, angular momenta and EoSs
\cite{AK1998,Benhar1999,KAA2001,Sotani2004a,Sotani2004b,Benhar2004}.
Thus, {\em gravitational wave asteroseismology} could enable us to
estimate the mass, radius and rotation rate of compact stars,
leading to the determination of the "best-candidate" high-density
EoS, which is still very uncertain. For this to happen, accurate
frequencies for different mode-sequences of rapidly rotating compact
objects have to be computed.

For slowly rotating stars, the frequencies of $f-$, $p-$ and $w-$
modes are still unaffected by rotation, and one can construct
approximate formulae in order to relate observed frequencies and
damping times of the various stellar modes to stellar parameters. For
example, for the fundamental oscillation ($l=2$) mode ($f$-mode) of
non-rotating stars one obtains \cite{AK1998}
\begin{eqnarray}
\sigma({\rm kHz})&\approx& 0.8+1.6 M_{1.4}^{1/2}R_{10}^{-3/2}
+ \delta_1 m{\bar \Omega} \\
\tau^{-1}({\rm secs}^{-1})&\approx&
M_{1.4}^3R_{10}^{-4}\left(22.9-14.7M_{1.4}R_{10}^{-1}\right)+
\delta_2 m {\bar \Omega}
\end{eqnarray}
where ${\bar \Omega}$ is the normalized rotation frequency of the
star, and $\delta_1$ and $\delta_2$ are constants estimated by
sampling data from various EOSs. The typical frequencies of NS
oscillation modes are larger than 1kHz. Since each type of mode is
sensitive to the physical conditions where the amplitude of the mode
is largest, the more oscillations modes can be identified through
gravitational waves, the better we will understand the detailed
internal structure of compact objects, such as the existence
of a possible superfluid state of matter \cite{Andersson01}.

If, on the other hand, some compact stars are born rapidly rotating
with moderate differential rotation, then their central densities
will be much smaller than the central density of a nonrotating star
of the same baryonic mass. Correspondingly, the typical axisymmetric
oscillation frequencies will be smaller than 1kHz, which is more
favorable for the sensitivity window of current interferometric
detectors \cite{SAF}. Indeed, axisymmetric simulations of rotating
core-collapse have shown that if a rapidly rotating NS is created,
then the dominant frequency of the core-bounce signal (originating
from the fundamental $l=2$ mode or the $l=2$ piece of the
fundamental quasi-radial mode) is in the range
600Hz-1kHz \cite{DFM2002b}.

If different types of signals are observed after core collapse,
such as both an axisymmetric core-bounce signal and a
nonaxisymmetric one-armed instability signal, with a time
separation of the order of 100ms, this would yield
invaluable information about the angular momentum distribution
in the proto-neutron stars.

\subsection*{ Acknowledgements} 
I am grateful to N. Andersson, N. Stergioulas, P. Lasky and A. Colaiuda for suggestions which improved the original manuscript.
This work was supported by the German Foundation (DFG) via SFB/TR7 and by the EU program ILIAS.


\vfill

\end{document}